\documentclass{article}
\usepackage[style = ieee]{biblatex}
\usepackage[utf8]{inputenc}
\usepackage{amsmath} % A math font library
\usepackage{float} % More figure positioning control
\usepackage[margin=1in]{geometry} % Margin width control
\usepackage{caption}
\usepackage{hanging}

\usepackage{amssymb}
\usepackage{graphics}
\usepackage{graphicx}
\usepackage{dirtytalk}
\usepackage{csquotes}
\usepackage[english]{babel}
\usepackage{amsthm}
\usepackage{listings}
\newtheorem{theorem}{Theorem}

\title{On the Stability of a Wormhole in the Maximally-Extended Reissner-Nordstr\"{o}m Solution}
\author{Ross DeMott \footnote{Please direct all correspondence to rdemott@mines.edu.}, \hspace{0.3 mm} Sam Major, and Alex Flournoy}
\date{\textit{Department of Physics, Colorado School of Mines, Golden, CO 80401, USA.} \\ \vspace{3 mm} \today}

\begin{document}
\maketitle

\begin{abstract}
\noindent We consider the stability of the maximally-extended Reissner-Nordstr\"{o}m solution in a Minkowski, de Sitter, or anti-de Sitter background.\footnote{By ``background," we mean the asymptotic behavior of the solution far from the singularity.}$^{,}$\footnote{Throughout this article, we will refer to these solutions as Reissner-Nordstr\"om or RN solutions, regardless of the background.} In a broad class of situations, prior work has shown that spherically symmetric perturbations from a massless scalar field cause the inner horizon of an RN black hole to become singular and collapse. Even if this is the case, it may still be possible for an observer to travel through the inner horizon before it fully collapses, thus violating strong cosmic censorship. In this work, we show that the collapse of the inner horizon and the occurrence of a singularity along the inner horizon are sufficient to prevent an observer from accessing the white hole regions and the parallel universe regions of the maximally extended RN space-time. Thus, if an observer passes through the inner horizon, they will inevitably hit the central singularity. Throughout this article, we use natural units where $c = G = 4 \pi \hspace{0.5 mm} \epsilon_{0} = 1$.
\end{abstract}

\section{Introduction} 
By definition, a black hole is an object that has an event horizon. The simplest black hole solution is the Schwarzschild solution, which possess an event horizon and no other horizons. However, if a black hole has charge and/or angular momentum, it exhibits an inner horizon as well as an event horizon. Such a black hole may be described by the Kerr-Newman solution. When the Kerr-Newman solution is maximally extended, the inner horizon forms the entrance to a wormhole, leading to a parallel universe \cite{PhysRev.174.1559}.
\vspace{2 mm}
\newline However, the Kerr-Newman solution is an idealization, as it contains no mass-energy except at the central singularity. To determine whether the inner horizon does actually form the entrance to a wormhole, we need to subject the black hole to perturbations. Moreover, we would like to prove or disprove the stability of the wormhole for a wide variety of perturbations. Eventually, we would like to consider arbitrary perturbations that exhibit no symmetry. For now, however, we will assume spherical symmetry to make the analysis more manageable. 
\vspace{2 mm}
\newline Any black hole with non-zero angular momentum lacks spherical symmetry, as the rotation axis specifies a preferred direction in space. Thus, we must start with a non-rotating black hole, and we must subject this black hole to spherically symmetric perturbations. When the angular momentum is zero, the Kerr-Newman solution reduces to the Reissner-Nordstr\"om solution. 
\vspace{2 mm}
\newline A Reissner-Nordstr\"{o}m (RN) black hole has both charge and mass, which are concentrated in a point-like singularity at $r = 0$. When the black hole is sub-extremal, this solution describes a black hole with two horizons: an outer event horizon and an inner Cauchy horizon. Between the event horizon and the inner horizon, all mass-energy is inexorably drawn inwards. This is equivalent to the statement that the radial coordinate $r$ is time-like in this region. However, inside the inner horizon, $r$ again becomes space-like, so it is possible for mass-energy to travel outwards from the singularity. The inner horizon is also a Cauchy horizon. In other words, given generic boundary conditions outside the inner horizon, it is generally impossible to find a unique solution for the space-time inside the inner horizon \cite{Dafermos2003, Jeffery1921}. This signals a breakdown of determinism.
\vspace{5 mm}
\newline In 1973, Simpson and Penrose demonstrated that the RN solution is unstable at the inner horizon \cite{Penrose1973}. Therefore, Penrose proposed that perturbations to the RN solution destroy the non-uniqueness of the unperturbed solution. This is known as the strong cosmic censorship (SCC) conjecture \cite{Dafermos2003-2}. More precisely, the SCC conjecture states that the instability at the inner horizon produces a singularity, preventing any observers from passing through it \cite{PhysRevLett.120.031103, PhysRevD.98.104007}. If true, this would imply that the region of space-time inside the inner horizon is unphysical. Hence, the entire physical space-time manifold would be uniquely specified by boundary conditions, and determinism would be restored.
\vspace{5 mm}
\newline Given boundary conditions outside the inner horizon, the inner horizon defines the boundary of the region of space-time where a unique solution for the metric can be found. Thus, it is possible to unambiguously describe the evolution of the inner horizon. Previous research has analyzed the behavior of the inner horizon in the presence of a spherically symmetric distribution of mass-energy. Dafermos showed that the metric can be extended continuously beyond the inner horizon, even in the presence of neutral scalar perturbations \cite{Dafermos2003, Dafermos2003-2}. Later, Costa et. al. numerically demonstrated an analogous result for near-extremal Reissner-Nordstr\"om (RN) black holes \cite{PhysRevLett.120.031103, PhysRevD.98.104007, EMS1, EMS2, EMS3, Costa2018}. 
\vspace{2 mm}
\newline In some scenarios, the in-falling mass-energy compresses near the inner horizon, creating a null curvature singularity called the mass-inflation singularity \cite{HISCOCK1981110, ChandHartle1982, Poisson1990, Brady1992, Brady1995, Ori1996, Burko1997, Ori1998, Burko1998, MTH2005, Ori2012, PhysRevD.19.413, PhysRevD.55.4860}. At first, these results may appear to contradict those mentioned in the previous paragraph. However, because the mass-inflation singularity is a weak singularity, the metric is continuous at the Cauchy horizon \cite{Ori1998, Ori2012}. Altogether, the results in Refs. \cite{Dafermos2003, Dafermos2003-2, PhysRevLett.120.031103, PhysRevD.98.104007, EMS1, EMS2, EMS3, Costa2018, ChandHartle1982, Poisson1990, Brady1992, Brady1995, Ori1996, Burko1997, Ori1998, Ori2012} suggest that the region inside the inner horizon may be accessible to observers. 
\vspace{2 mm}
\newline Before we proceed, it is worth noting that the mass-inflation instability is not generic to all relativistic theories of gravity. In some modified theories of gravity, such as $f\left(R\right)$ gravity, the mass-inflation instability is absent \cite{doi:10.1142/S0219887822500736, doi:10.1142/S0219887822500281, SHAMIR2021312, USMAN2022101691, doi:10.1142/S0217751X21502031}. In these theories, models of stellar collapse do not result in any singularities forming outside the physical singularity. Additionally, there is some evidence that quantum effects may dampen or eliminate the mass-inflation singularity \cite{Tang_2019}. We will leave the question of wormhole stability in the absence of a mass-inflation singularity for future work. In this paper, we assume that a mass-inflation instability occurs at the inner horizon. We also assume that it is possible for an observer to pass through the inner horizon. Under this assumption, we seek to determine the ultimate fate of such an observer.
\vspace{5 mm}
\newline Given that the space-time inside the Cauchy horizon cannot be uniquely specified, it is not possible to describe an observer's trajectory inside the Cauchy horizon with certainty. However, if we assume that space-time can still be treated classically inside the inner horizon, it is possible to make concrete statements about the observer's ultimate fate. In the unperturbed, maximally extended RN space-time, the region between the inner horizon and the inner anti-horizon is a wormhole (see Figure \ref{fig:Unperturbed R-N Penrose Diagram}), which leads to a white hole and a parallel universe. (In fact, the maximally extended RN solution contains an infinite number of parallel universes.) Using the Raychaudhuri equation \cite{PhysRevD.98.084029}, we show that this wormhole collapses in the presence of spherically symmetric perturbations satisfying the null energy condition. More precisely, we show that, if a mass-inflation instability occurs on the Cauchy horizon, any time-like or light-like observer who passes through the Cauchy horizon of a perturbed RN black hole will inevitably hit the central singularity. Thus, we conclude that the parallel universes of the unperturbed RN space-time are unphysical.

\section{Coordinate Systems for Reissner-Nordstr\"om Black Holes}
Let us consider a black hole with mass $M$ and charge $Q$, which is subjected to massless scalar perturbations. We assume that the metric and all fields are spherically-symmetric. In spherical coordinates, a general spherically symmetric metric takes the form
\begin{equation} \label{generalsphericallysymmetricmetricrt}
ds^{2} = -f\left(r, t\right) \hspace{0.5 mm} dt^{2} + f\left(r, t\right)^{-1} \hspace{0.5 mm} dr^{2} + r^{2} \hspace{0.5 mm} d \Omega^{2} .
\end{equation} 
Let $\Lambda$ be the cosmological constant. In the absence of perturbations, the RN metric has \cite{Brady1992, Chambers:1997ef}
\begin{equation}
f\left(r\right) = 1 - \frac{2 M}{r} + \frac{Q^{2}}{r^{2}} - \frac{1}{3} \hspace{0.5 mm} \Lambda \hspace{0.5 mm} r^{2} .
\end{equation}
Let $\mathcal{U}$ and $\mathcal{V}$ be general double null coordinates. In such coordinates, we may rewrite Eqn. \ref{generalsphericallysymmetricmetricrt} as
\begin{equation} \label{generalsphericallysymmetricmetricuv}
ds^{2} = -F\left(\mathcal{U}, \mathcal{V}\right) \hspace{0.5 mm} d\mathcal{U} \hspace{0.5 mm} d\mathcal{V} + r^{2}\left(\mathcal{U}, \mathcal{V}\right) \hspace{0.5 mm} d\Omega^{2} .
\end{equation}
A sub-extremal RN black hole has two horizons: an event horizon and an inner Cauchy horizon. In the absence of perturbations, such a space-time has the following Penrose diagram:
\begin{figure}[H]
    \centering
    \includegraphics[scale=0.45]{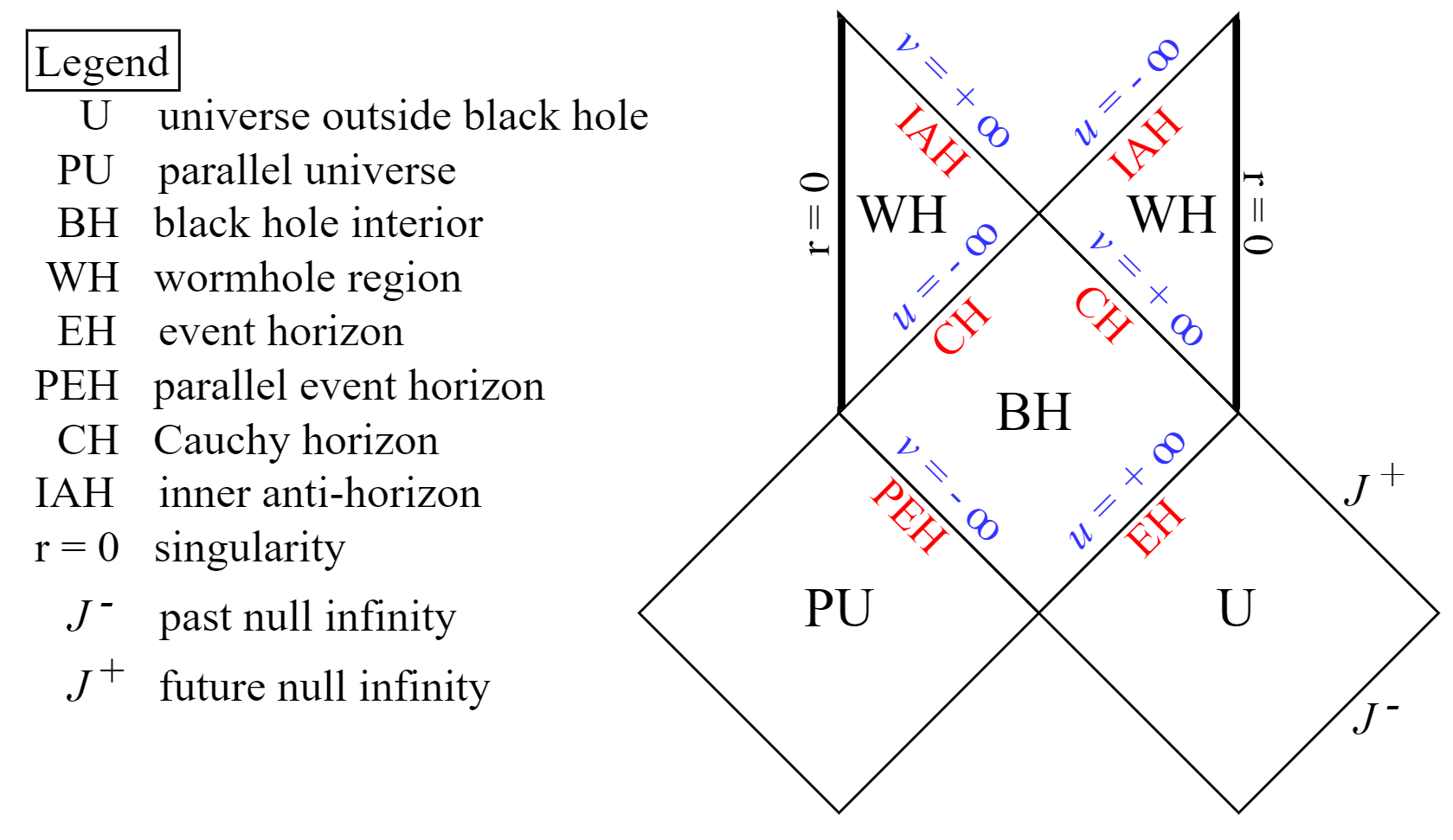}
        \caption{This is a section of the maximally-extended Penrose diagram for an unperturbed, sub-extremal RN black hole. This diagram uses Eddington-Finkelstein (EF) coordinates $u$ and $v$, which we will define shortly.}
    \label{fig:Unperturbed R-N Penrose Diagram}
\end{figure}
\noindent In Figure \ref{fig:Unperturbed R-N Penrose Diagram}, the bottom right square represents our universe, while the bottom left square represents a parallel universe. Because the parallel universe is inaccessible to any observer traveling from our universe, we may regard it as unphysical. With this in mind, all perturbations must enter the black hole through the event horizon in our universe. 

\subsection{Eddington-Finkelstein Double-Null Coordinates} \label{EFcoordinates}
Now, we define the Eddington-Finkelstein (EF) double-null coordinates $u$ and $v$ for an unperturbed RN black hole \cite{Ori2012}. Technically, there are two sets of EF coordinates: one for the region inside the inner horizon, and one for the region outside the inner horizon. Let $r_{-}$ denote the radius of the inner horizon for an unperturbed RN black hole. Let $\kappa$ represent the gravitational acceleration at the inner horizon. We may write $\kappa$ as \cite{Ori1998, Ori2012}
\begin{equation}
\kappa = -\frac{1}{2} \hspace{0.5 mm} \left. \frac{d f}{dr} \right|_{r = r_{-}} .
\end{equation}
We define the tortoise coordinate $r^{*}$ to satisfy the relation
\begin{equation}
\frac{d r}{d r^{*}} = f\left(r\right) .
\end{equation}
Because $f\left(r\right) > 0$ inside the inner horizon and $f\left(r\right) < 0$ outside the inner horizon, $\kappa$ is positive. Close to $r = r_{-}$, we may approximate $f\left(r\right)$ as
\begin{equation}
f\left(r\right) \approx -2 \kappa \hspace{0.5 mm} \left(r - r_{-}\right) .
\end{equation}
Let $A$ be an arbitrary real constant. Close to the inner horizon, we may write the solution $r\left(r^{*}\right)$ as
\begin{equation}
\label{rofrstar}
r\left(r^{*}\right) = r_{-} +  A \hspace{0.5 mm} \exp\left[-2 \kappa \hspace{0.5 mm} r^{*}\right] .
\end{equation}
Now, we want to invert Eqn. \ref{rofrstar} to obtain $r^{*}\left(r\right)$. To accomplish this, we must consider the regions outside and inside the inner horizon separately.
\vspace{2 mm}
\newline First, we consider the region outside the inner horizon ($r > r_{-}$). In order for $r^{*}$ to be real, the constant $A$ must be positive. For simplicity, we choose $A = 1$. Thus, we may rewrite Eqn. \ref{rofrstar} as 
\begin{equation}
\label{exprstar}
\exp\left[-2 \kappa \hspace{0.5 mm} r^{*}\left(r\right)\right] = r - r_{-} ,
\end{equation}
\begin{equation}
r^{*}\left(r\right) = - \frac{1}{2 \kappa} \hspace{0.5 mm} \ln\left(r - r_{-}\right) .
\end{equation}
Next, we consider the region inside the inner horizon ($r < r_{-}$). In order for $r^{*}$ to be real, the constant $A$ must be negative. For simplicity, we choose $A = -1$. Thus, we may rewrite Eqn. \ref{rofrstar} as
\begin{equation}
\exp\left[-2 \kappa \hspace{0.5 mm} r^{*}\left(r\right)\right] = r_{-} - r ,
\end{equation}
\begin{equation}
r^{*}\left(r\right) = - \frac{1}{2 \kappa} \hspace{0.5 mm} \ln\left(r_{-} - r\right) .
\end{equation}
Combining Equations 10 and 12, we may write $r^{*}$ as
\begin{equation}
r^{*} = -\frac{1}{2 \kappa} \hspace{0.5 mm} \ln \left|r - r_{-}\right| .
\end{equation}
Finally, we introduce the Eddington-Finkelstein double-null coordinates $u$ and $v$, defined as
\begin{equation} \label{EFu}
u = t - r^{*} ,
\end{equation}
\begin{equation} \label{EFv}
v = t + r^{*} .
\end{equation}
Just outside the inner horizon, we may approximate $F\left(u, v\right)$ (defined in Eqn. \ref{generalsphericallysymmetricmetricuv}) as
\begin{equation} \label{outerF}
F\left(u, v\right) \approx -2 \kappa \hspace{1 mm} e^{\kappa \hspace{0.5 mm} \left(u - v\right)} .
\end{equation}
Just inside the inner horizon, we may approximate $F\left(u, v\right)$ as
\begin{equation} \label{innerF}
F\left(u, v\right) \approx 2 \kappa \hspace{1 mm} e^{\kappa \hspace{0.5 mm} \left(u - v\right)} .
\end{equation}
In the outer EF double-null coordinate system, the inner horizon lies at the limits $v \to \infty$ and $u \to -\infty$. On the $u = -\infty$ section of the Cauchy horizon, a severe singularity occurs \cite{Brady1995, Burko1997, Burko1998, Ori2012}. Because of this, objects cannot pass through this section of the Cauchy horizon. Therefore, the wormhole region beyond the $u = -\infty$ section of the Cauchy horizon is unphysical. From here on, we shall primarily focus on the $v = + \infty$ section of the Cauchy horizon.

\subsection{Kruskal-Szekeres Transformation on the $v$ Coordinate for an Unperturbed RN Black Hole} \label{StandardCoordinateSystem}
The Eddington-Finkelstein double-null coordinate system is the simplest double null coordinate system to derive from standard spherical coordinates. From here on, it will be convenient to use the Eddington-Finkelstein coordinate $u$ as one of our double null coordinates. However, we note that there are two separate $v$ coordinates: one for the region outside the inner horizon and another for the region inside the inner horizon. Fortunately, it is possible to glue these two $v$ coordinates together by combining both of them into a single Kruskal-Szekeres coordinate $V$. By convention, we choose the inner horizon to be at $V = 0$. We choose $V$ to be positive outside the inner horizon and negative inside the inner horizon. Outside the inner horizon, we define $V$ as
\begin{equation} \label{KSout}
V\left(v\right) = e^{- \kappa v} .
\end{equation}
Inside the inner horizon, we define $V$ as
\begin{equation} \label{KSin}
V\left(v\right) = - e^{- \kappa v} .
\end{equation}
Since $u$ and $V$ are both null coordinates, the metric still takes the form
\begin{equation}
ds^{2} = -F\left(u, V\right) \hspace{0.5 mm} du \hspace{0.5 mm} dV + r^{2}\left(u, V\right) \hspace{0.5 mm} d\Omega^{2} .
\end{equation}
From Equations \ref{outerF} and \ref{innerF}, we know how $F\left(u, v\right)$ behaves close to the inner horizon in Eddington-Finkelstein coordinates. With basic calculus, we may use the known expression for $F\left(u, v\right)$ to find $F\left(u, V\right)$. Close to the inner horizon (both outside and inside), we may approximate $F\left(u, V\right)$ as
\begin{equation} \label{KrusSzekF}
F\left(u, V\right) \approx 2 \hspace{0.5 mm} e^{\kappa u} .
\end{equation}
We derived Eqn. \ref{KrusSzekF} using an unperturbed RN black hole space-time. Thus, one might be tempted to conclude that Eqn. \ref{KrusSzekF} only applies to an unperturbed RN black hole. However, in a perturbed space-time, we may assume that $F\left(u, V\right)$ is positive, finite, and continuous everywhere along the Cauchy horizon (except at the physical singularity) \cite{Ori1998, EMS1, EMS2, EMS3}. Thus, even for a perturbed black hole, we may perform a gauge transformation such that $F\left(u, V\right)$ matches Eqn. \ref{KrusSzekF} close to the inner horizon.

\section{Space-Time Dynamics and the Raychaudhuri Equation in General Double Null Coordinates}
In this section, we describe the dynamics of an arbitrary spherically symmetric metric coupled to a massless Klein-Gordon scalar field $\phi$. We also calculate the Raychaudhuri scalar $\Theta$ and describe its dynamics. These equations hold in arbitrary double null coordinates.

\subsection{Einstein Field Equations and Dynamics of the Scalar Field}
Let us consider a general spherically symmetric metric (Eqn. \ref{generalsphericallysymmetricmetricuv}). Let $\phi$ be a massless Klein-Gordon scalar field, and let $\Lambda$ be the cosmological constant. We may write the Einstein field equations as
\begin{equation} \label{EFELambdaruv}
r_{, u v} = -\frac{r_{, u} \hspace{0.5 mm} r_{, v}}{r} - \frac{F}{4 r} \hspace{0.5 mm} \left(1 - \frac{Q^{2}}{r^{2}}\right) + \frac{\Lambda}{4} \hspace{0.5mm} r \hspace{0.5 mm} F ,    
\end{equation}
\begin{equation} \label{EFELambdaFuv}
F_{, u v} = \frac{F_{, u} \hspace{0.5 mm} F_{, v}}{F} + \frac{2F}{r^{2}} \hspace{0.5 mm} r_{, u} \hspace{0.5 mm} r_{, v} + \frac{F^{2}}{2 r^{2}} \hspace{0.5 mm} \left(1 - \frac{2 Q^{2}}{r^{2}}\right) - 2 F \hspace{0.5 mm} \phi_{, u} \hspace{0.5 mm} \phi_{, v} + \Lambda \hspace{0.5 mm} F^{2} ,
\end{equation}
\begin{equation} \label{EFELambdaruu}
r_{, u u} - \left(\ln F\right)_{, u} \hspace{0.5 mm} r_{, u} + r \hspace{0.5 mm} \left(\phi_{, u}\right)^{2} = 0 ,
\end{equation}
\begin{equation} \label{EFELambdarvv}
r_{, v v} - \left(\ln F\right)_{, v} \hspace{0.5 mm} r_{, v} + r \hspace{0.5 mm} \left(\phi_{, v}\right)^{2} = 0 .
\end{equation}
The derivation of Eqns. \ref{EFELambdaruv}-\ref{EFELambdarvv} may be found in Appendix B. The scalar field $\phi$ satisfies the massless Klein-Gordon equation \cite{Ori1998}:
\begin{equation} \label{masslessKG}
\phi_{, u v} + \frac{1}{r} \hspace{0.5 mm} \left(r_{, u} \hspace{0.5 mm} \phi_{, v} + r_{, v} \hspace{0.5 mm} \phi_{, u}\right) = 0 .
\end{equation}
We assume that $\phi$ is a spherically symmetric function, so there is no angular dependence. At the Cauchy horizon, infalling mass-energy creates a singularity called the mass inflation instability. In effect, the mass-energy ``piles up" at the Cauchy horizon \cite{HISCOCK1981110, ChandHartle1982, Poisson1990, Brady1992, Brady1995, Ori1996, Burko1997, Ori1998, Burko1998, MTH2005, Ori2012}. Below, we have included a figure to illustrate this effect with massless radiation.
\begin{figure}[H]
    \centering
    \includegraphics[scale=0.50]{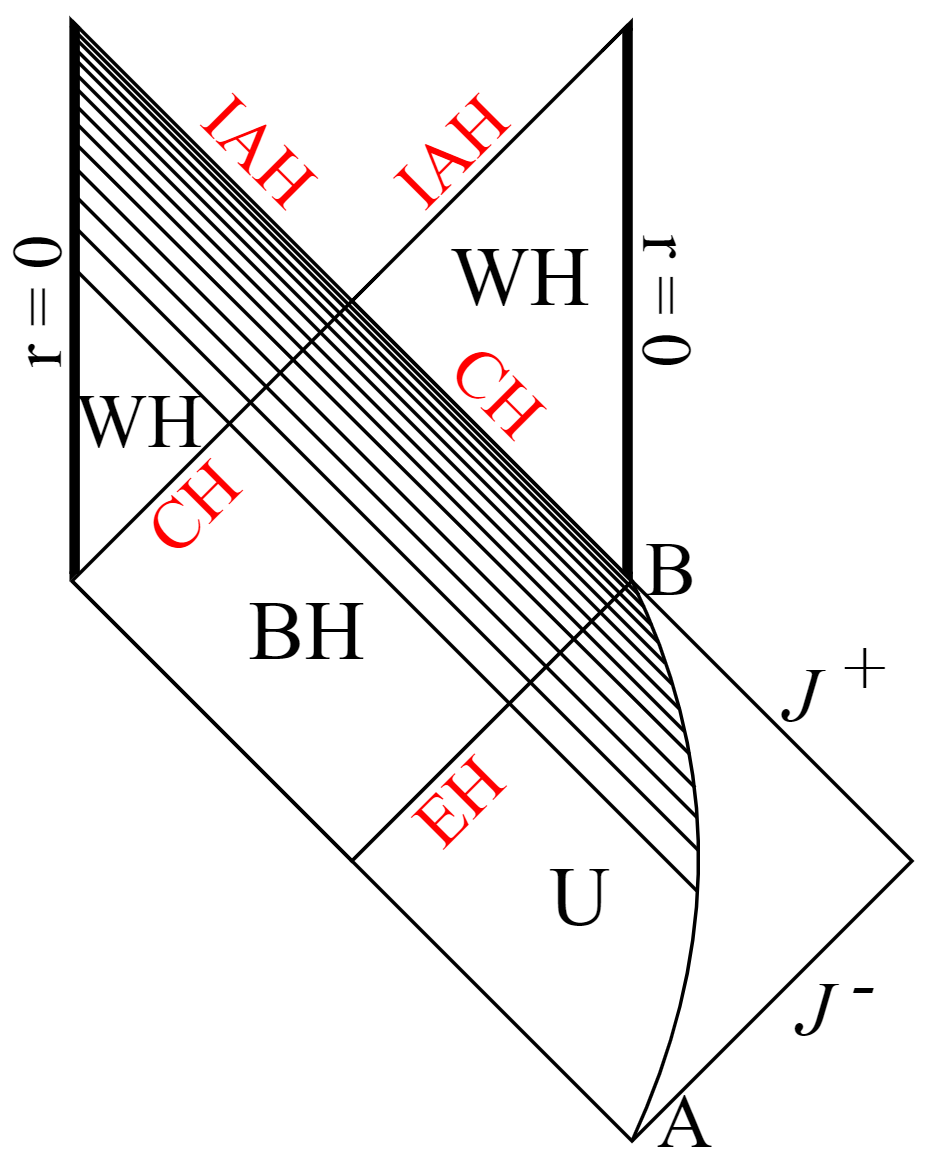}
        \caption{A Penrose diagram illustrating the formation of the mass inflation singularity at the Cauchy horizon. Light is emitted from the curve $AB$ in the region U. As it gets closer to the Cauchy horizon, the light waves become compressed and blue-shifted. This causes the energy of the light to increase without bound, creating a singularity.}
    \label{fig:Mass Inflation Penrose Diagram}
\end{figure}
\noindent As discussed in the introduction, the mass inflation instability is a weak singularity, which means that it is possible for an observer to pass through the Cauchy horizon \cite{Ori1998, Ori2012}.

\subsection{Raychaudhuri Equation} \label{raychaudhurisubsection}
For a more thorough derivation of all the statements in this subsection, please see Appendix A. Let us define the vector fields $\ell^{\mu}$ and $n^{\mu}$ as
\begin{equation} \label{elldefmain}
\ell^{\mu} = \begin{cases}
F\left(u, v\right)^{-1}, & \textrm{if } \mu = v \\
0, & \textrm{otherwise}
\end{cases} ,
\end{equation}
\begin{equation}
n^{\mu} = \begin{cases}
2, & \textrm{if } \mu = u \\
0, & \textrm{otherwise}
\end{cases} .
\end{equation}
Let $\Theta$ be the Raychaudhuri expansion scalar in the $v$ direction. We may write $\Theta$ as
\begin{equation} \label{thetaexpression}
\Theta = 2 \hspace{0.5 mm} F\left(u, v\right)^{-1} \hspace{0.5 mm} r\left(u, v\right)^{-1} \hspace{0.5 mm} \partial_{v} r\left(u, v\right) .
\end{equation}
The scalar $\Theta$ obeys the Raychaudhuri equation
\begin{equation} \label{raychaudhurieqn}
\ell^{\mu} \hspace{0.5 mm} \partial_{\mu} \Theta = -\frac{1}{2} \Theta^{2} - R_{\mu \nu} \hspace{0.5 mm} \ell^{\mu} \ell^{\nu} .
\end{equation}
The total stress-energy tensor (Eqn. \ref{Ttotal}) satisfies the null energy condition
\begin{equation} \label{nullenergycondn}
T_{\mu \nu} \hspace{0.5 mm} \ell^{\mu} \hspace{0.5 mm} \ell^{\nu} \geq 0 .
\end{equation}
Plugging Eqn. \ref{nullenergycondn} into Eqn. \ref{raychaudhurieqn}, we find that
\begin{equation}
\partial_{v} \Theta \leq 0 .
\end{equation}

\section{Sufficient Conditions for a Wormhole to be Unstable} \label{ProofsSec}
In this section, we specify several conditions that are sufficient for a spherically-symmetric wormhole, formed from a sub-extremal RN black hole solution, to be unstable. We use the coordinate system from Subsection \ref{StandardCoordinateSystem}, with $u$ an Eddington-Finkelstein coordinate and $V$ a Kruskal-Szekeres coordinate. We assume that the black hole solution has a Cauchy horizon.

\subsection{Regularity Conditions} \label{regconds}
We assume that the functions $F\left(u, V\right)$ and $r\left(u, V\right)$ are finite, positive, and continuous for all $u$ and $V$, except at the physical singularity \cite{Ori1998}. At all points $u$, $V$ not at the physical singularity or on the Cauchy horizon, we assume that $F\left(u, V\right)$ and $r\left(u, V\right)$ are twice continuously differentiable. We also assume that $r_{, u}\left(u, V\right)$ is continuously differentiable for all $u$ and $V$ (including $V = 0$), except at the physical singularity. 
\vspace{2 mm}
\newline These regularity conditions are fairly generic, and we have not found any situation (relevant to this paper) in which they have been shown to be violated. Thus, we will assume that these conditions are true in all the situations we consider in this paper.

\subsection{Situation-Dependent Conditions} \label{specificconds}
In addition to the regularity conditions, we have two conditions that cannot be assumed to hold generally. Thus, we will have to determine whether these conditions hold in each of the situations we consider.
\vspace{2 mm}
\newline We assume that there is at least one point on the Cauchy horizon, which we call $u_{1}$, such that 
\begin{equation} \label{rurcond}
r_{, u}\left(u = u_{1}, V = 0\right) > \kappa \hspace{0.5 mm} r\left(u = u_{1}, V = 0\right) .
\end{equation}
We also assume that there exists another point $u_{2}$ on the Cauchy horizon, which satisfies $u_{2} < u_{1}$, such that the following limit holds from both sides:
\begin{equation} \label{rvderivlimit}
\lim_{V \to \hspace{0.5 mm} 0} r_{, V}\left(u = u_{2}, V\right) = + \infty .
\end{equation}

\subsection{Theorem 1: Collapse of the Cauchy Horizon to $r = 0$}
\begin{theorem} \label{thmcol}
Let the assumptions described in Subsection \ref{regconds} be true. Additionally, let Eqn. \ref{rurcond} be true. Then, for all points $u < u_{1}$ on the Cauchy horizon (except at the physical singularity),
\begin{equation}
r_{, u}\left(u, V = 0\right) > 0 .
\end{equation}
Additionally, there exists a finite value $u_{\textrm{min}}$ such that
\begin{equation}
r\left(u = u_{\textrm{min}}, V = 0\right) = 0 .
\end{equation}
\end{theorem}
\noindent \textbf{Proof.}
Using Eqn. \ref{KrusSzekF}, we may rewrite Eqn. \ref{EFELambdaruu} as
\begin{equation} \label{thmcolEFEruu}
r_{, u u} - \kappa \hspace{0.5 mm} r_{, u} = - r \hspace{0.5 mm} \left(\phi_{, u}\right)^{2} .
\end{equation}
Note that $\kappa$ is non-negative. Using the product rule, we may rewrite Eqn. \ref{thmcolEFEruu} as
\begin{equation} \label{thmcolEFEruurewritten}
\frac{d}{d u} \left(e^{- \kappa u} \hspace{0.5 mm} r_{, u}\right) = - e^{- \kappa u} \hspace{0.5 mm} r \hspace{0.5 mm} \left(\phi_{, u}\right)^{2} .
\end{equation}
From Eqn. \ref{thmcolEFEruurewritten}, it is easy to see that
\begin{equation} \label{thmcolEFEzero}
\frac{d}{d u} \left(e^{- \kappa u} \hspace{0.5 mm} r_{, u}\right) \leq 0.
\end{equation}
For any value $u < u_{1}$, Eqn. \ref{thmcolEFEzero} implies that
\begin{equation} \label{thmcolru}
e^{- \kappa u} \hspace{0.5 mm} r_{, u}\left(u, V = 0\right) \geq e^{- \kappa u_{1}} \hspace{0.5 mm} r_{, u}\left(u = u_{1}, V = 0\right) .
\end{equation}
Rearranging Eqn. \ref{thmcolru}, we obtain
\begin{equation} \label{thmcolru2}
r_{, u}\left(u, V = 0\right) \geq e^{\kappa \hspace{0.5 mm} \left(u - u_{1}\right)} \hspace{0.5 mm} r_{, u}\left(u = u_{1}, V = 0\right) .
\end{equation}
Eqn. \ref{thmcolru2} implies the first part of Theorem \ref{thmcol}. For all points $u < u_{1}$ on the Cauchy horizon (except at the physical singularity),
\begin{equation}
r_{, u}\left(u, V = 0\right) > 0 .
\end{equation}
Next, we integrate both sides of Eqn. \ref{thmcolru2}, which yields
\begin{align}
r\left(u, V = 0\right) & = r\left(u = u_{1}, V = 0\right) - \int_{u}^{u_{1}} r_{, u}\left(u^{\prime}, V = 0\right) \hspace{0.5 mm} \mathrm{d}u^{\prime} \\
& \leq r\left(u = u_{1}, V = 0\right) - r_{, u}\left(u = u_{1}, V = 0\right) \hspace{0.5 mm} \int_{u}^{u_{1}} e^{\kappa \hspace{0.5 mm} \left(u^{\prime} - u_{1}\right)} \hspace{0.5 mm} \mathrm{d}u^{\prime} \\
& = r\left(u = u_{1}, V = 0\right) - \frac{r_{, u}\left(u = u_{1}, V = 0\right)}{\kappa} + \frac{r_{, u}\left(u = u_{1}, V = 0\right)}{\kappa} \hspace{0.5 mm} e^{\kappa \hspace{0.5 mm} \left(u - u_{1}\right)} . \label{thmcolalignlast}
\end{align}
From the assumptions in Subsection \ref{specificconds}, we know that
\begin{equation}
r\left(u = u_{1}, V = 0\right) - \frac{r_{, u}\left(u = u_{1}, V = 0\right)}{\kappa} < 0 .
\end{equation}
As $u$ decreases, the third term in Eqn. \ref{thmcolalignlast} becomes arbitrarily small. Thus, there exists a finite value $u_{\textrm{min}}$ such that
\begin{equation}
r\left(u = u_{\textrm{min}}, V = 0\right) = 0 .
\end{equation}

\subsection{Theorem 2: Divergence of $r_{, V}$ at the Cauchy Horizon}
\begin{theorem}
\label{thmdivrv}
Let the assumptions described in Subsections \ref{regconds} and \ref{specificconds} be true. Then, for all $u \in \left(u_{\textrm{min}}, u_{2}\right]$, the following statement is true:
\begin{equation}
\lim_{V \to \hspace{0.5 mm} 0} r_{, V}\left(u, V\right) = + \infty .
\end{equation}
\end{theorem}
\noindent \textbf{Proof.}
We proceed via proof by contradiction. Let us assume that there is some point $u_{\textrm{test}} \in \left(u_{\textrm{min}}, u_{2}\right)$ such that
\begin{equation} \label{assumcontradictthmdivrv}
\lim_{V \to 0} r_{, V}\left(u = u_{\textrm{test}}, V\right) \neq +\infty .
\end{equation}
In any interval $V \in \left(-\epsilon, 0\right) \cup \left(0, \epsilon\right)$, there will be at least one line $V = V_{1}$ such that
\begin{equation} \label{V1firstline}
r_{, V}\left(u = u_{\textrm{test}}, V = V_{1}\right) < r_{, V}\left(u = u_{2}, V = V_{1}\right) .
\end{equation}
Because $V_{1} \neq 0$, we know that $r_{, V}\left(u, V = V_{1}\right)$ is continuous for all $u > u_{\textrm{min}}$. From Equation \ref{EFELambdaruv}, we have the following expression for $r_{, u V}$:
\begin{equation}
r_{, u v} = -\frac{r_{, u} \hspace{0.5 mm} r_{, v}}{r} - \frac{F}{4 r} \hspace{0.5 mm} \left(1 - \frac{Q^{2}}{r^{2}}\right) + \frac{\Lambda}{4} \hspace{0.5mm} r \hspace{0.5 mm} F .
\end{equation}
Since Equation \ref{EFELambdaruv} holds for any double null coordinate system, we have replaced $v$ with the Kruskal-Szekeres coordinate $V$. 
\vspace{2 mm}
\newline The functions $F\left(u = u_{2}, V\right)$, $r\left(u = u_{2}, V\right)$, and $r_{, u}\left(u = u_{2}, V\right)$ are finite, positive, and continuous. By contrast, $r_{, V}\left(u = u_{2}, V\right)$ grows without bound as $V \to 0$. Therefore, if we choose $V_{1}$ to be sufficiently close to zero, we can make $r_{, u V}\left(u = u_{2}, V = V_{1}\right)$ negative.
\vspace{2 mm}
\newline Because $r_{, u V}$ is continuous along the line $V = V_{1}$, there must be some interval around $u = u_{2}$ (meaning $u_{2}$ is not on the boundary) where $r_{, u V}\left(u, V = V_{1}\right)$ is negative. Let $\mathcal{J}$ be the largest interval, without any gaps, such that $u_{2} \in \mathcal{J}$ and such that, for all $u \in \mathcal{J}$,
\begin{equation}
r_{, u V}\left(u, V = V_{1}\right) < 0 .
\end{equation}
Let $u = u_{3}$ be the infimum of the set $\mathcal{J}$. If $u_{3} = u_{\textrm{min}}$, then for all points $u \in \left(u_{\textrm{min}}, u_{2}\right)$,
\begin{equation} \label{contradictionuinf=unin}
r_{, V}\left(u, V = V_{1}\right) > r_{, V}\left(u_{2}, V = V_{1}\right) .
\end{equation}
Clearly, Eqn. \ref{contradictionuinf=unin} contradicts our finding in Eqn. \ref{V1firstline}. Thus, if $u_{3} = u_{\textrm{min}}$, Theorem \ref{thmdivrv} is proved. 
\vspace{2 mm}
\newline Now, let us assume that $u_{3} > u_{\textrm{min}}$. Because $r_{, u V}\left(u, V = V_{1}\right)$ is continuous, we know that 
\begin{equation}
r_{, u V}\left(u = u_{3}, V = V_{1}\right) = 0 .
\end{equation}
For all $u \in \left(u_{3}, u_{2}\right)$, we know that
\begin{equation} \label{uu2inequality}
r_{, V}\left(u, V = V_{1}\right) > r_{, V}\left(u = u_{2}, V = V_{1}\right) .
\end{equation}
Because $r_{, V}\left(u, V = V_{1}\right)$ is continuous along the line $V = V_{1}$, Eqn. \ref{uu2inequality} implies that
\begin{equation} \label{u3u2inequality}
r_{, V}\left(u = u_{3}, V = V_{1}\right) > r_{, V}\left(u = u_{2}, V = V_{1}\right) .
\end{equation}
On the closed interval $\mathcal{I} = \left[u_{3}, u_{2}\right]$, the functions $F\left(u, V = V_{1}\right)$ and $r\left(u, V = V_{1}\right)$ are continuous and positive. If we choose $V_{1}$ to be sufficiently close to $V = 0$, then $r_{, u}\left(u, V = V_{1}\right)$ will also be continuous and positive on $\mathcal{I}$ (by Theorem \ref{thmcol}).
\vspace{2 mm}
\newline According to the extreme value theorem \cite{trench_2003}, $F\left(u, V = V_{1}\right)$, $r\left(u, V = V_{1}\right)$, and $r_{, u}\left(u, V = V_{1}\right)$ all have positive lower and upper bounds on $\mathcal{I}$. Therefore, we may choose $V_{1}$ such that, for all $u \in \mathcal{I}$:
\begin{equation} \label{importantinequality}
r_{, V}\left(u = u_{2}, V = V_{1}\right) > - \frac{F\left(u, V = V_{1}\right)}{4 r_{, u}\left(u, V = V_{1}\right)} + \frac{Q^{2} \hspace{0.5 mm} F\left(u, V = V_{1}\right)}{4 r\left(u, V = V_{1}\right)^{2} \hspace{0.5 mm} r_{, u}\left(u, V = V_{1}\right)} .    
\end{equation}
Using Equation \ref{u3u2inequality}, we may rewrite Eqn. \ref{importantinequality} as
\begin{equation} \label{secondimportantinequality}
r_{, V}\left(u = u_{3}, V = V_{1}\right) > - \frac{F\left(u, V = V_{1}\right)}{4 r_{, u}\left(u, V = V_{1}\right)} + \frac{r^{2}_{Q} \hspace{0.5 mm} F\left(u, V = V_{1}\right)}{4 r\left(u, V = V_{1}\right)^{2} \hspace{0.5 mm} r_{, u}\left(u, V = V_{1}\right)} .  
\end{equation}
Plugging Eqn. \ref{secondimportantinequality} into Eqn. \ref{EFELambdaruv}, we find that 
\begin{equation} \label{secondimporineqintoEFE}
r_{, u V}\left(u = u_{3}, V = V_{1}\right) < 0 .
\end{equation}
However, Eqn. \ref{secondimporineqintoEFE} conflicts with our earlier finding that 
\begin{equation}
r_{, u V}\left(u = u_{3}, V = V_{1}\right) = 0 .
\end{equation}
Thus, if there is some point $u_{\textrm{test}} \in \left(u_{\textrm{min}}, u_{2}\right)$ such that 
\begin{equation}
\lim_{V \to 0} r_{, V}\left(u = u_{2}, V\right) \neq \infty ,
\end{equation}
we obtain a contradiction. Therefore, for all $u \in \left(u_{\textrm{min}}, u_{2}\right]$, we know that
\begin{equation}
\lim_{V \to 0} r_{, V}\left(u, V\right) = +\infty .
\end{equation}

\subsection{Theorem 3: Positivity of $r_{, V}$ Inside the Cauchy Horizon}
\begin{theorem}
\label{thmposrv}
Let the assumptions described in Subsections \ref{regconds} and \ref{specificconds} be true. Then,
\begin{equation}
r_{, V}\left(u, V\right) > 0 \textrm{ for all } u, v \in \left(u_{\textrm{min}}, u_{2}\right] \times \left(-\infty, 0\right) .
\end{equation}
\end{theorem}
\noindent \textbf{Proof.}
Let us consider an arbitrary value $u_{\textrm{test}} \in \left(u_{min}, u_{2}\right]$. From Theorem \ref{thmdivrv}, we have
\begin{equation}
\lim_{V \to \hspace{0.5 mm} 0} r_{, V}\left(u = u_{\textrm{test}}, V\right) = + \infty .
\end{equation}
Therefore, there exists some real number $\epsilon > 0$ such that 
\begin{equation}
r_{, V}\left(u = u_{\textrm{test}}, V\right) > 0 \textrm{ for all } V \in \left(-\epsilon, 0\right) \cup \left(0, \epsilon\right) .
\end{equation}
Recall Equation \ref{thetaexpression} for the Raychaudhuri scalar $\Theta$:
\begin{equation}
\Theta = 2 \hspace{0.5 mm} F\left(u, v\right)^{-1} \hspace{0.5 mm} r\left(u, v\right)^{-1} \hspace{0.5 mm} \partial_{V} r\left(u, v\right) .
\end{equation}
For any $u$ and $V$ in the physical space-time, if $r_{, V}\left(u, V\right) > 0$, then $\Theta\left(u, V\right) > 0$. Therefore, there exists some real number $\epsilon > 0$ such that
\begin{equation}
\Theta\left(u = u_{\textrm{test}}, V\right) > 0 \textrm{ for all } V \in \left(-\epsilon, 0\right) \cup \left(0, \epsilon\right) .
\end{equation}
From Subsection \ref{raychaudhurisubsection}, we know that $\partial_{V} \Theta \leq 0$. Therefore, for any $V < 0$, we know that 
\begin{equation}
\Theta\left(u_{\textrm{test}}, V\right) > 0 .
\end{equation}
This implies that, for all $V < 0$,
\begin{equation}
r_{, V}\left(u_{\textrm{test}}, V\right) > 0 .
\end{equation}
Since $u_{\textrm{test}}$ is an arbitrary element of the interval $\left(u_{min}, u_{2}\right]$, we find that 
\begin{equation}
r_{, V}\left(u, V\right) > 0 \textrm{ for all } u, V \in \left(u_{\textrm{min}}, u_{2}\right] \times \left(-\infty, 0\right) .
\end{equation}

\subsection{Application to Wormholes}
Let us assume that the conditions in Subsections \ref{regconds} and \ref{specificconds} are true. Furthermore, let us assume that 
\begin{equation}
r_{, u}\left(u = u_{1}, V = 0\right) \geq \kappa \hspace{0.5 mm} r\left(u = u_{1}, V = 0\right) .
\end{equation}
Then, from Theorem \ref{thmcol}, we know that $r\left(u, V = 0\right)$ either reaches zero at a finite value $u = u_{\textrm{min}}$, or
\begin{equation}
\lim_{u \to -\infty} r\left(u, V = 0\right) = 0 .
\end{equation}
According to Theorem \ref{thmposrv},
\begin{equation}
r_{, V}\left(u, V\right) > 0 \textrm{ for all } u, V \in \left(u_{\textrm{min}}, u_{1}\right] \times \left(-\infty, 0\right) .
\end{equation}
Therefore, for any $u_{0} \in \left(u_{\textrm{min}}, u_{1}\right)$ and any $V_{0} < 0$,
\begin{equation}
r\left(u = u_{0}, V = V_{0}\right) < r\left(u = u_{0}, V = 0\right) .
\end{equation}
Because
\begin{equation}
\lim_{u \to u_{\textrm{min}}} r\left(u, V = 0\right) \to 0 ,
\end{equation}
for any $V_{0} < 0$, there exists some $u_{\textrm{zero}} \geq u_{\textrm{min}}$ such that
\begin{equation}
\lim_{u \to u_{\textrm{zero}}} r\left(u, V = V_{0}\right) = 0 .
\end{equation}
Therefore, any object that passes through the Cauchy horizon will inevitably hit the physical singularity $r = 0$. Thus, the Penrose diagram for our space-time looks similar to the one below.
\begin{figure}[H]
    \centering
    \includegraphics[scale=0.50]{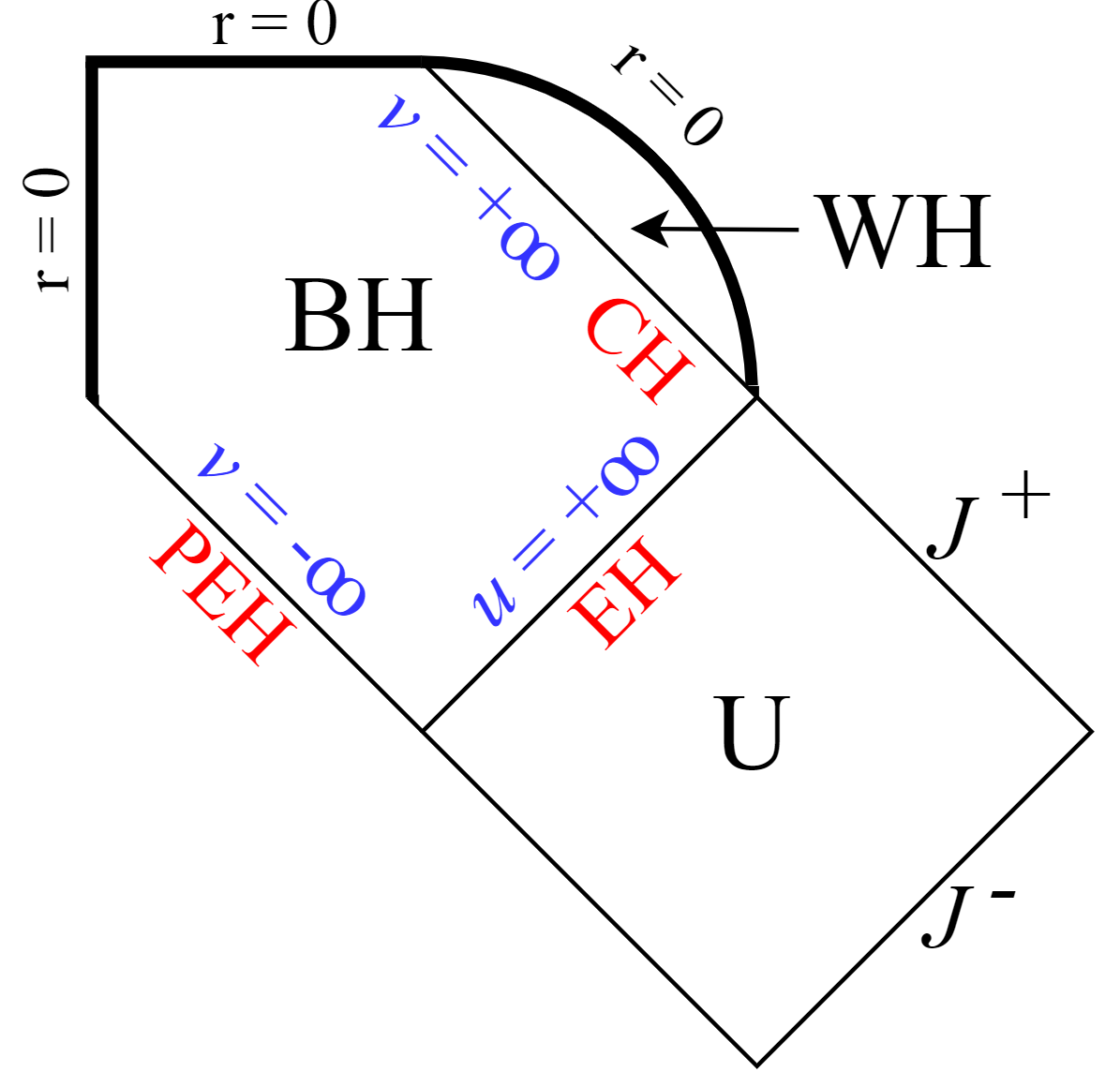}
        \caption{Penrose diagram for a perturbed Reissner-Nordstr\"{o}m black hole. The values of the double-null coordinates \textit{u} and \textit{v} at the horizons are shown in blue. Note that $u$ and $v$ are both Eddington-Finkelstein null coordinates.}
    \label{fig:Perturbed R-N Penrose Diagram}
\end{figure}

\section{Wormhole Stability for a Sub-Extremal RN Black Hole}
In this section, we determine under what conditions a sub-extremal RN black hole satisfies the assumptions in Subsection \ref{specificconds}. We adopt a coordinate system similar to that of Subsection \ref{StandardCoordinateSystem}, with $u$ an Eddington-Finkelstein coordinate and $V$ a Kruskal-Szekeres coordinate.

\subsection{Zero Cosmological Constant ($\Lambda = 0$)}
Let us consider a sub-extremal RN black hole in the absence of a cosmological constant. The radius of the Cauchy horizon decreases monotonically as $u$ decreases \cite{Poisson1990, Brady1995, Burko1997, Ori1998}. At a finite value $u = u_{\textrm{min}}$, the Cauchy horizon intersects with the physical singularity at $r = 0$ \cite{Burko1997, Ori2012}. 
\vspace{2 mm}
\newline According to Ref. \cite{Ori1998}, the derivative $r_{, V}\left(u, V\right)$ approaches $+ \infty$ as $V \to 0$ from the outside of the Cauchy horizon. Because $r_{, V}\left(u, V\right) \to +\infty$ as $V \to 0$ outside the Cauchy horizon, it is reasonable to assume that there is at least one line $u = u_{1}$ where the same limit holds from both sides of the Cauchy horizon.
\vspace{2 mm}
\newline For $\Lambda = 0$, all the conditions specified in Subsection \ref{specificconds} are satisfied. Thus, a wormhole formed from a Reissner-Nordstr\"om black hole, in the absence of a cosmological constant, is unstable. Note that this result holds for any sub-extremal RN black hole, as long as $\Lambda = 0$.

\subsection{Positive Cosmological Constant ($\Lambda > 0$)}
Let us consider a black hole embedded in a background de Sitter space with cosmological constant $\Lambda$. On the event horizon of the black hole, we assume that the actual metric matches the RN metric and that the first derivatives of the actual metric match the first derivatives of the RN metric \cite{EMS1, EMS2, EMS3}.
\vspace{2 mm}
\newline Let $r_{+}$ be the event horizon radius of an unperturbed RN black hole with mass $M$ and charge $Q$. Let $r_{-}$ be the Cauchy horizon radius of such a black hole. Now, we define the ratios $\sigma$ and $\Upsilon$ as \cite{EMS1, EMS2, EMS3}
\begin{equation}
\sigma = \frac{r_{+}}{r_{-}} ,
\end{equation}
\begin{equation}
\Upsilon = \frac{\Lambda \hspace{0.5 mm} r_{-}^{2}}{3} .
\end{equation}
In terms of $\sigma$ and $\Upsilon$, we may define the ratio $\rho$ as \cite{EMS1, EMS2, EMS3}
\begin{equation}
\rho = \sigma^{2} \hspace{0.5 mm} \frac{1 - \Upsilon \hspace{0.5 mm} \left(\sigma^{2} + 2 \sigma + 3\right)}{1 - \Upsilon \hspace{0.5 mm} \left(3 \sigma^{2} + 2 \sigma + 1\right)} .
\end{equation}
In the limit of a Schwarzschild-de Sitter black hole ($r_{-} = 0$), the ratio $\rho$ approaches $+ \infty$. In the limit of an extremal RN black hole ($r_{-} = r_{+}$), the ratio $\rho$ approaches unity \cite{EMS1, EMS2, EMS3}. Based on the ratio $\rho$, we divide RN black holes into two classes. If $\rho > 2$, we call the black hole a near-Schwarzschild RN black hole. If $\rho \leq 2$, we call the black hole a near-extremal RN black hole.

\subsubsection{Near-Schwarzschild RN Black Hole}
Let us assume that $\rho > 2$, so that we are working with a near-Schwarzschild RN black hole. Let $V_{0}$ be a finite number; we specify initial data for the field $\phi$ on the surface $V = V_{0}$. For a broad variety of initial conditions (more specifically, if $r_{, u}\left(u, V_{0}\right)$ fails to decay sufficiently quickly close to the event horizon), a mass-inflation instability will emerge on the Cauchy horizon of the black hole \cite{EMS1, EMS2, EMS3}. 
\vspace{2 mm}
\newline The radius of the Cauchy horizon satisifes \cite{EMS2, EMS3}
\begin{equation} \label{CauchyHorizonRLimit}
\lim_{u \to +\infty} r\left(u, V = 0\right) = r_{-} .
\end{equation}
Therefore, there exists some constant $U_{1}$ such that, for all $u > U_{1}$,
\begin{equation} \label{U1eqn}
r\left(u, V = 0\right) > 0 .
\end{equation}
There exists a constant $U_{2}$ such that, for all $u > U_{2}$ \cite{EMS2, EMS3},
\begin{equation} \label{U2eqn}
r_{, u}\left(u, V = 0\right) > 0 .
\end{equation}
On the event horizon, the first derivatives of the actual metric match the first derivatives of the RN metric. Thus, the radius of the Cauchy horizon should satisfy \cite{EMS2, EMS3}
\begin{equation}
\lim_{u \to + \infty} r_{, u}\left(u, V = 0\right) = 0 .
\end{equation}
Thus, there exists some value $U_{3}$ such that, for all $u > U_{3}$, 
\begin{equation} \label{U3eqn}
r_{, u}\left(u, V = 0\right) < + \infty. 
\end{equation}
Let us define the renormalized Hawking mass $\omega\left(u, v\right)$ as \cite{EMS1, EMS2, EMS3}
\begin{equation} \label{omegadef}
\omega\left(u, V\right) = \frac{e^{2}}{2} \hspace{0.5 mm} r\left(u, V\right)^{-1} + \frac{1}{2} \hspace{0.5 mm} r\left(u, V\right) - \frac{\Lambda}{6} \hspace{0.5 mm} r\left(u, V\right)^{3} + 2 \hspace{0.5 mm} F\left(u, V\right)^{-2} r\left(u, V\right) \hspace{0.5 mm} r_{, u}\left(u, V\right) \hspace{0.5 mm} r_{, V}\left(u, V\right) .
\end{equation}
Note that $\omega\left(u, V\right)$ is a gauge invariant quantity. As discussed above, a mass-inflation instability occurs for generic perturbations in this space-time. Thus, there exists a constant $U_{4}$ such that, for all $u > U_{4}$ \cite{EMS3},
\begin{equation} \label{U4eqn}
\lim_{V \to 0} \omega\left(u, V\right) = + \infty .
\end{equation}
Let us select a constant $U$ that is larger than $U_{1}$, $U_{2}$, $U_{3}$, and $U_{4}$. Combining Eqns. \ref{U1eqn}, \ref{U2eqn}, \ref{U3eqn}, \ref{omegadef}, and \ref{U4eqn}, we find that, for all $u > U$,
\begin{equation}
\lim_{V \to 0} r_{, V}\left(u, V\right) = + \infty .
\end{equation}
All the conditions specified in Subsections \ref{regconds} and \ref{specificconds} are satisfied. Thus, under general initial conditions, a wormhole formed from a near-Schwarzschild Reissner-Nordstr\"om black hole is unstable if $\Lambda > 0$.

\subsubsection{Near-Extremal RN Black Hole}
Let us assume that $\rho < 2$, so that we are working with a near-extremal RN black hole. In general, it is not possible to guarantee that a mass-inflation instability will occur on the Cauchy horizon of such a black hole. In fact, there exist examples of near-extremal RN black holes that fail to develop a mass-inflation instability under a broad range of initial conditions \cite{EMS1, EMS2, EMS3}.
\vspace{2 mm}
\newline If no mass-inflation instability occurs, then the function $\omega\left(u, v\right)$ remains finite at the Cauchy horizon. Thus, it is not possible for all the conditions in Subsections \ref{regconds} and \ref{specificconds} to be fulfilled. This implies that a wormhole formed from a near-extremal RN black hole may be stable if $\Lambda > 0$. 

\subsection{Negative Cosmological Constant ($\Lambda < 0$)}
In the presence of a negative cosmological constant, the Cauchy horizon collapses to $r = 0$ \cite{PhysRevD.51.5517, PhysRevD.54.7898}. However, at all points on the Cauchy horizon where $r \neq 0$, there is no singularity \cite{PhysRevD.51.5517}. Therefore, $r_{, V}\left(u, V = 0 \right)$ remains finite except at the physical singularity ($r = 0$). This violates the conditions of Subsection \ref{specificconds}. Therefore, the wormhole that occurs inside an RN black hole may be stable if $\Lambda < 0$.

\section{Wormhole Stability for an Extremal RN Black Hole}
An extremal RN black hole has only one horizon, which is simultaneously an event horizon and a Cauchy horizon \cite{Aretakis1, Aretakis2, MassiveExtremeRN}. Let $r_{h}$ denote the radius of this horizon. Let us define the tortoise coordinate $r^{*}$ as \cite{MassiveExtremeRN, Murata_2013}
\begin{equation} \label{exactextremetortoisecoordinate}
r^{*}\left(r\right) = r - \hspace{0.5 mm} r_{h} + 2 r_{h} \hspace{0.5 mm} \ln\left(\frac{\lvert r - r_{h} \rvert}{r_{h}}\right) - \frac{r_{h}^{2}}{r - r_{h}} .
\end{equation}
We define the EF double null coordinates $u$ and $v$ as \cite{MassiveExtremeRN, Murata_2013}
\begin{equation} \label{extremeEFu}
u = t - r^{*} ,
\end{equation}
\begin{equation} \label{extremeEFv}
v = t + r^{*} .
\end{equation}
In these EF coordinates, the horizon corresponds to the limit $u \to +\infty$ \cite{MassiveExtremeRN}. For a sub-extremal black hole, the Cauchy horizon corresponds to the limit $v \to + \infty$. To rectify this disparity, we simply swap the definitions of the coordinates $u$ and $v$.
\begin{equation} \label{extremeEFuswapped}
u = t + r^{*} ,
\end{equation}
\begin{equation} \label{extremeEFvswapped}
v = t - r^{*} .
\end{equation}
In order to obtain meaningful results for physics at the horizon, we must define a new coordinate $V$ that is regular at the horizon. We may do this via the following Kruskal-Szekeres transformation \cite{Murata_2013, MassiveExtremeRN}:
\begin{equation}
v = -2 r^{*} \hspace{0.5 mm} \left(r_{h} + V\right) .
\end{equation}
In this new coordinate system, the horizon lies at $V = 0$. The region outside the horizon corresponds to $V > 0$, while the region inside the horizon corresponds to $V < 0$. Close to the horizon, we have the following approximation for $F\left(u, V\right)$ \cite{Murata_2013}:
\begin{equation}
F\left(u, V\right) \approx 2 .
\end{equation}
As in the sub-extremal case, we assume that the conditions in Subsections \ref{regconds} and \ref{specificconds} are fulfilled. For an extremal black hole, the surface gravity $\kappa$ is zero \cite{Murata_2013}. Thus, we may rewrite Eqn. \ref{rurcond} as
\begin{equation}
r_{, u}\left(u = u_{1}, V = 0\right) > 0 \textrm{ for some } u_{1} \in \mathbb{R} .
\end{equation}
We may rewrite Eqn. \ref{EFELambdaruu} as
\begin{equation} \label{EFELambdaruuextremal}
r_{, u u} = - r \hspace{0.5 mm} \left(\phi_{, u}\right)^{2} .
\end{equation}
Clearly, Eqn. \ref{EFELambdaruuextremal} implies that $r_{, u u}$ is non-positive everywhere along the Cauchy horizon. Thus, for all $u < u_{1}$,
\begin{equation} \label{ruuruu1ineqextremal}
r_{, u}\left(u, V = 0\right) > r_{, u}\left(u = u_{1}, V = 0\right) .
\end{equation}
From Eqn. \ref{ruuruu1ineqextremal}, we see that there exists a finite value $u_{\textrm{min}}$ such that
\begin{equation}
r\left(u = u_{\textrm{min}}, V = 0\right) = 0 .
\end{equation}
Having proved Theorem \ref{thmcol}, one may prove Theorems \ref{thmdivrv} and \ref{thmposrv} for an extremal RN black hole in the same way as for a sub-extremal RN black hole. Thus, a wormhole formed from an extremal RN black hole collapses. It is impossible for an observer to access the parallel universe regions of the maximally-extended extremal RN space-time.

\section{Summary and Future Work}
Under the assumptions of Subsections \ref{regconds} and \ref{specificconds}, we have shown that a Reissner-Nordstr\"{o}m wormhole collapses in the presence of spherically symmetric perturbations from a massless scalar field. In this article, we have held firmly to the assumption of spherical symmetry. In the future, it would be interesting to investigate the stability of non-spherically symmetric wormholes, such as those that arise from the Kerr-Newman solution. Additionally, it would be interesting to incorporate non-spherically symmetric distributions of mass-energy as perturbations.
\vspace{2 mm}
\newline \noindent As the wormhole collapses, it will eventually become sufficiently small that quantum effects become important. At this scale, the classical analysis performed in this paper is no longer valid. Hence, it is unknown whether the wormhole will continue to collapse or not. Future work is needed to assess the impact of quantum effects on the collapse of the wormhole.

\section*{Appendix A: Computation of the Expansion Scalar and the Raychaudhuri Equation in General Double Null Coordinates}
Let $K$ be a scalar field, and let $\ell^{\mu}$ and $n^{\mu}$ be null vector fields. Let $\nabla_{\nu}$ denote a covariant derivative. The null vector fields must satisfy the following conditions:
\begin{equation} \label{RayCond1}
\ell^{\mu} \hspace{0.5 mm} n_{\mu} = -1 ,
\end{equation}
\begin{equation} \label{RayCond2}
\ell^{\nu} \hspace{0.5 mm} \nabla_{\nu} \ell^{\mu} = K \hspace{0.5 mm} \ell^{\mu} .
\end{equation}
We may define the expansion scalar $\Theta$ as
\begin{equation} \label{Thetadef}
\Theta = \nabla_{\mu} \ell^{\mu} - K .
\end{equation}
Next, let us define the transverse metric $h_{\mu \nu}$ as
\begin{equation}
h_{\mu \nu} = g_{\mu \nu} + \ell_{\mu} \hspace{0.5 mm} n_{\nu} + \ell_{\nu} \hspace{0.5 mm} n_{\mu} .
\end{equation}
Finally, let us define the quantities $\sigma_{\mu \nu}$ and $\omega_{\mu \nu}$ as
\begin{equation}
\sigma_{\mu \nu} = \frac{1}{2} \left(\nabla^{\sigma} \ell^{\rho}\right) \hspace{0.5 mm} \left(h_{\mu \rho} \hspace{0.5 mm} h_{\nu \sigma} + h_{\nu \rho} \hspace{0.5 mm} h_{\mu \sigma}\right) - \frac{1}{2} h_{\mu \nu} \hspace{0.5 mm} \Theta ,
\end{equation}
\begin{equation}
\omega_{\mu \nu} = \frac{1}{2} \left(\nabla^{\sigma} \ell^{\rho}\right) \hspace{0.5 mm} \left(h_{\mu \rho} \hspace{0.5 mm} h_{\nu \sigma} - h_{\nu \rho} \hspace{0.5 mm} h_{\mu \sigma}\right) .
\end{equation}
Let us make the definitions $\sigma^{2} := \sigma_{\mu \nu} \hspace{0.5 mm} \sigma^{\mu \nu}$ and $\omega^{2} := \omega_{\mu \nu} \hspace{0.5 mm} \omega^{\mu \nu}$). The Raychaudhuri equation takes the following form \cite{PhysRevD.98.084029}: 
\begin{equation}
\ell^{\mu} \hspace{0.5 mm} \partial_{\mu} \Theta = -\frac{1}{2} \Theta^{2} - \sigma^{2} + \omega^{2} - R_{\mu \nu} \hspace{0.5 mm} \ell^{\mu} \ell^{\nu} + K \Theta .
\end{equation}
We use the index $u$, not to be confused with $\mu$, to denote the component of a vector in the positive $u$ direction. We use the index $v$, not to be confused with $\nu$, to denote the component of a vector in the positive $v$ direction. We are free to choose the vector fields $\ell^{\mu}$ and $n^{\mu}$, as long as they satisfy Equations \ref{RayCond1} and \ref{RayCond2}. Therefore, we select the following expressions for $\ell^{\mu}$ and $n^{\mu}$:
\begin{equation} \label{elldef}
\ell^{\mu} = \begin{cases}
F\left(u, v\right)^{-1}, & \textrm{if } \mu = v \\
0, & \textrm{otherwise}
\end{cases} ,
\end{equation}
\begin{equation} \label{ndef}
n^{\mu} = \begin{cases}
2, & \textrm{if } \mu = u \\
0, & \textrm{otherwise}
\end{cases} .
\end{equation}
To avoid confusion with the Latin indices $u$ and $v$, let us replace the Greek indices $\mu$ and $\nu$ in Equation \ref{RayCond2} with $\alpha$ and $\beta$. We may expand $\ell^{\beta} \hspace{0.5 mm} \nabla_{\beta} \ell^{\alpha}$ as follows
\begin{align}
\ell^{\beta} \hspace{0.5 mm} \nabla_{\beta} \ell^{\alpha} & = \ell^{\beta} \hspace{0.5 mm} \left(\partial_{\beta} \ell^{\alpha} + \Gamma^{\alpha}_{\beta \gamma} \hspace{0.5 mm} \ell^{\gamma}\right) \\
& = \ell^{v} \hspace{0.5 mm} \partial_{v} \ell^{\alpha} + \Gamma^{\alpha}_{v v} \hspace{0.5 mm} \left(\ell^{v}\right)^{2} .
\end{align}
We may write the Christoffel symbols $\Gamma^{\alpha}_{v v}$ as follows:
\begin{align}
\Gamma^{\alpha}_{v v} & = g^{\alpha \beta} \hspace{0.5 mm} \partial_{v} g_{v \beta} \\
& = \begin{cases}
F\left(u, v\right)^{-1} \hspace{0.5 mm} \partial_{v} F\left(u, v\right) & \textrm{if } \alpha = v \\
0 & \textrm{otherwise}
\end{cases} .
\end{align}
When $\alpha \neq v$, we find that $\ell^{\beta} \hspace{0.5 mm} \nabla_{\beta} \ell^{\alpha} = 0$. Therefore, the only non-trivial component of Equation \ref{RayCond2} occurs when $\alpha = v$. We may write this component as
\begin{equation}
\ell^{v} \hspace{0.5 mm} \partial_{v} \ell^{v} + \Gamma^{v}_{v v} \hspace{0.5 mm} \left(\ell^{v}\right)^{2} = K \hspace{0.5 mm} \ell^{v} .
\end{equation}
Thus, we may write the scalar $K$ as
\begin{align}
K & = \partial_{v} \ell^{v} + \Gamma^{v}_{v v} \hspace{0.5 mm} \ell^{v} \\
& = -\left(F\left(u, v\right)\right)^{-2} \hspace{0.5 mm} \partial_{v} F\left(u, v\right) + F\left(u, v\right)^{-2} \hspace{0.5 mm} \partial_{v} F\left(u, v\right) \\
& = 0 .
\end{align}
Using Equation \ref{Thetadef}, we may write $\Theta$ as
\begin{align}
\Theta & = \nabla_{\alpha} \ell^{\alpha} - K \\
& = \partial_{\alpha} \ell^{\alpha} + \Gamma^{\alpha}_{\alpha \gamma} \ell^{\gamma} \\
& = \partial_{v} \ell^{v} + \Gamma^{\alpha}_{\alpha v} \ell^{v} .
\end{align}
The index $\alpha$ is summed, since it does not denote a specific coordinate direction. We may write the quantity $\Gamma^{\alpha}_{\alpha v}$ as
\begin{align}
\Gamma^{\alpha}_{\alpha v} & = \frac{1}{2} \hspace{0.5 mm} g^{\alpha \beta} \hspace{0.5 mm} \left(\partial_{v} g_{\alpha \beta} + \partial_{\alpha} g_{v \beta} - \partial_{\beta} g_{v \alpha}\right) \\
& = g^{u v} \hspace{0.5 mm} \partial_{v} g_{u v} + \frac{1}{2} \hspace{0.5 mm} g^{\theta \theta} \hspace{0.5 mm} \partial_{v} g_{\theta \theta} + \frac{1}{2} \hspace{0.5 mm} g^{\phi \phi} \hspace{0.5 mm} \partial_{v} g_{\phi \phi} \\
& = F\left(u, v\right)^{-1} \hspace{0.5 mm} \partial_{v} F\left(u, v\right) + 2 \hspace{0.5 mm} r\left(u, v\right)^{-1} \hspace{0.5 mm} \partial_{v} r\left(u, v\right) .
\end{align}
Therefore, we may write $\Theta$ as
\begin{equation}
\Theta = 2 \hspace{0.5 mm} F\left(u, v\right)^{-1} \hspace{0.5 mm} r\left(u, v\right)^{-1} \hspace{0.5 mm} \partial_{v} r\left(u, v\right) .
\end{equation}
Recall the expression for the transverse metric $h_{\mu \nu}$ ($\mu$ and $\nu$ are Greek indices):
\begin{equation}
h_{\mu \nu} = g_{\mu \nu} + \ell_{\mu} \hspace{0.5 mm} n_{\nu} + \ell_{\nu} \hspace{0.5 mm} n_{\mu} .
\end{equation}
With $\ell^{\mu}$ and $n^{\mu}$ defined by Equations $\ref{elldef}$ and $\ref{ndef}$, it is easy to show that $h_{\mu \nu}$ is non-zero if and only $\mu = \nu = \theta$ or $\mu = \nu = \phi$. Hence, we have the following expression for $h_{\mu \nu}$:
\begin{equation}
h_{\mu \nu} = \begin{cases}
g_{\theta \theta} & \textrm{if } \mu = \nu = \theta \\
g_{\phi \phi} & \textrm{if } \mu = \nu = \phi \\
0 & \textrm{otherwise}
\end{cases} .
\end{equation}
Now, we seek to prove that $\sigma^{2} = 0$ and $\omega^{2} = 0$. The tensors $\sigma_{\mu \nu}$ and $\omega_{\mu \nu}$ are zero unless both $\mu$ and $\nu$ are elements of the set $\left\{\theta, \phi\right\}$. The tensor $\sigma_{\mu \nu}$ is symmetric in $\mu$ and $\nu$, while the tensor $\omega_{\mu \nu}$ is anti-symmetric in $\mu$ and $\nu$. Because $\omega_{\mu \nu}$ is anti-symmetric in $\mu$ and $\nu$, $\omega_{\theta \theta}$ and $\omega_{\phi \phi}$ must both be zero.
\vspace{2 mm}
\newline Below, we prove that $\omega_{\theta \phi} = 0$:
\begin{align}
\omega_{\theta \phi} & = \frac{1}{2} \hspace{0.5 mm} \left(\nabla^{\sigma} \ell^{\rho}\right) \hspace{0.5 mm} \left(h_{\theta \rho} \hspace{0.5 mm} h_{\phi \sigma} - h_{\phi \rho} \hspace{0.5 mm} h_{\theta \sigma}\right) \\
& = \frac{1}{2} \hspace{0.5 mm} \left(\nabla^{\phi} \ell^{\theta}\right) \hspace{0.5 mm} h_{\theta \theta} \hspace{0.5 mm} h_{\phi \phi} - \frac{1}{2} \hspace{0.5 mm} \left(\nabla^{\theta} \ell^{\phi}\right) \hspace{0.5 mm} h_{\theta \theta} \hspace{0.5 mm} h_{\phi \phi} \\
& = \frac{1}{2} \hspace{0.5 mm} h_{\theta \theta} \hspace{0.5 mm} h_{\phi \phi} \hspace{0.5 mm} \left(\nabla^{\phi} \ell^{\theta} - \nabla^{\theta} \ell^{\phi}\right) \\
& = \frac{1}{2} \hspace{0.5 mm} h_{\theta \theta} \hspace{0.5 mm} h_{\phi \phi} \hspace{0.5 mm} g^{\theta \theta} \hspace{0.5 mm} g^{\phi \phi} \hspace{0.5 mm} \left(\nabla_{\phi} \ell_{\theta} - \nabla_{\theta} \ell_{\phi}\right) \\
& = \frac{1}{2} \hspace{0.5 mm} \left(\partial_{\phi} \ell_{\theta} - \partial_{\theta} \ell_{\phi}\right) \\
& = 0 .
\end{align}
Since all components of $\omega_{\mu \nu}$ are zero, the scalar $\omega^{2}$ is also zero. Below, we prove that $\sigma_{\theta \theta} = 0$:
\begin{align}
\sigma_{\theta \theta} & = \left(\nabla^{\theta} \ell^{\theta}\right) \hspace{0.5 mm} \left(h_{\theta \theta}\right)^{2} - \frac{1}{2} \hspace{0.5 mm} h_{\theta \theta} \hspace{0.5 mm} \Theta \\
& = \nabla_{\theta} \ell_{\theta} - \frac{1}{2} \hspace{0.5 mm} g_{\theta \theta} \hspace{0.5 mm} \Theta \\
& = - \Gamma_{\alpha \theta \theta} \hspace{0.5 mm} \ell^{\alpha} - \frac{1}{2} \hspace{0.5 mm} g_{\theta \theta} \hspace{0.5 mm} \Theta \\
& = \frac{1}{2} \hspace{0.5 mm} \ell^{v} \hspace{0.5 mm} \partial_{v} g_{\theta \theta} - \frac{1}{2} \hspace{0.5 mm} g_{\theta \theta} \hspace{0.5 mm} \Theta \\
& = F\left(u, v\right)^{-1} \hspace{0.5 mm} r\left(u, v\right) \hspace{0.5 mm} \partial_{v} r\left(u, v\right) - F\left(u, v\right)^{-1} \hspace{0.5 mm} r\left(u, v\right) \hspace{0.5 mm} \partial_{v} r\left(u, v\right) \\
& = 0 .
\end{align}
A similar calculation shows that $\sigma_{\phi \phi} = 0$. Finally, we prove that $\sigma_{\theta \phi} = 0$:
\begin{align}
\sigma_{\theta \phi} & = \frac{1}{2} \hspace{0.5 mm} \left(\nabla^{\phi} \ell^{\theta}\right) \hspace{0.5 mm} h_{\theta \theta} \hspace{0.5 mm} h_{\phi \phi} + \frac{1}{2} \hspace{0.5 mm} \left(\nabla^{\theta} \ell^{\phi}\right) \hspace{0.5 mm} h_{\theta \theta} \hspace{0.5 mm} h_{\phi \phi} - \frac{1}{2} \hspace{0.5 mm} h_{\theta \phi} \hspace{0.5 mm} \Theta \\
& = \frac{1}{2} \hspace{0.5 mm} \left(\nabla_{\theta} \ell_{\phi} + \nabla_{\phi} \ell_{\theta}\right) \\
& = - \Gamma_{\alpha \theta \phi} \hspace{0.5 mm} \ell^{\alpha} \\
& = 0 .
\end{align}
Because all components of $\sigma_{\mu \nu}$ are zero, we have proven that $\sigma^{2} = 0$. Thus, the Raychaudhuri scalar $\Theta$ obeys the equation
\begin{equation}
\ell^{\mu} \hspace{0.5 mm} \partial_{\mu} \Theta = -\frac{1}{2} \Theta^{2} - R_{\mu \nu} \hspace{0.5 mm} \ell^{\mu} \ell^{\nu} .
\end{equation}
Let us define $T$ by $T = g^{ab} \hspace{0.5 mm} T_{ab}$. Using the trace-reversed Einstein field equations, we may write $R_{a b}$ as
\begin{equation}
R_{ab} - \Lambda \hspace{0.5 mm} g_{a b} = 8 \pi \hspace{0.5 mm} T_{ab} - 4 \pi \hspace{0.5 mm} T \hspace{0.5 mm} g_{ab} .
\end{equation}
Next, we substitute the trace-reversed EFE into the Raychaudhuri equation. Because $\ell^{\mu}$ is a null vector, the terms involving $T$ and $\Lambda$ disappear. Hence, we have the following equation:
\begin{equation}
\ell^{\mu} \hspace{0.5 mm} \partial_{\mu} \Theta = -\frac{1}{2} \Theta^{2} - 8 \pi \hspace{0.5 mm} T_{\mu \nu} \hspace{0.5 mm} \ell^{\mu} \ell^{\nu} .
\end{equation}
Now, we impose the null energy condition $T_{\mu \nu} \hspace{0.5 mm} \ell^{\mu} \hspace{0.5 mm} \ell^{\nu} \geq 0$. With this condition, it is clear that $\ell^{\mu} \hspace{0.5 mm} \partial_{\mu} \Theta$ is strictly non-positive. In symbols,
\begin{equation}
\ell^{\mu} \hspace{0.5 mm} \partial_{\mu} \Theta \leq 0 .
\end{equation}
Because the only non-zero component of $\ell^{\mu}$ is $\ell^{v}$, which is positive, we know that $\partial_{v} \Theta \leq 0$.

\section*{Appendix B: Spherically-Symmetric Einstein Field Equations with a Cosmological Constant in Double Null Coordinates}
With $\Lambda = 0$, the Einstein field equations take the following form \cite{Ori1998}:
\begin{equation} \label{maineineqn}
r_{, u v} = -\frac{r_{, u} \hspace{0.5 mm} r_{, v}}{r} - \frac{F}{4 r} \hspace{0.5 mm} \left(1 - \frac{Q^{2}}{r^{2}}\right) ,    
\end{equation}
\begin{equation} \label{Feqn}
F_{, u v} = \frac{F_{, u} \hspace{0.5 mm} F_{, v}}{F} + \frac{2F}{r^{2}} \hspace{0.5 mm} r_{, u} \hspace{0.5 mm} r_{, v} + \frac{F^{2}}{2 r^{2}} \hspace{0.5 mm} \left(1 - \frac{2 Q^{2}}{r^{2}}\right) - 2 F \hspace{0.5 mm} \phi_{, u} \hspace{0.5 mm} \phi_{, v} ,
\end{equation}
\begin{equation} \label{reqnu}
r_{, u u} - \left(\ln F\right)_{, u} \hspace{0.5 mm} r_{, u} + r \hspace{0.5 mm} \left(\phi_{, u}\right)^{2} = 0 ,
\end{equation}
\begin{equation} \label{reqnv}
r_{, v v} - \left(\ln F\right)_{, v} \hspace{0.5 mm} r_{, v} + r \hspace{0.5 mm} \left(\phi_{, v}\right)^{2} = 0 .
\end{equation}
The scalar field $\phi$ has an associated stress-energy tensor $T_{\alpha \beta}^{s}$ \cite{Ori1998}:
\begin{equation} \label{Tscalar}
T_{\alpha \beta}^{s} = \frac{1}{8 \pi} \hspace{0.5 mm} \left(\phi_{, \alpha} \hspace{0.5 mm} \phi_{, \beta} - \frac{1}{2} \hspace{0.5 mm} g_{\alpha \beta} \hspace{0.5 mm} g^{\gamma \delta} \hspace{0.5 mm} \phi_{, \gamma} \hspace{0.5 mm} \phi_{, \delta}\right) .
\end{equation}
The charge $Q$ generates a Coulomb field around the black hole. This electric field also has an associated stress-energy tensor $T^{\textrm{EM}}_{\alpha \beta}$. We may write the non-zero components of $T^{\textrm{EM}}_{\alpha \beta}$ as \cite{Ori1998}
\begin{equation}
T^{\textrm{EM}}_{u v} = T^{\textrm{EM}}_{v u} = \frac{Q^{2}}{16 \pi} \hspace{0.5 mm} F\left(u, v\right) \hspace{0.5 mm} r\left(u, v\right)^{-4} ,
\end{equation}
\begin{equation}
T^{\textrm{EM}}_{\theta \theta} = \frac{Q^{2}}{8 \pi} \hspace{0.5 mm} r\left(u, v\right)^{-2} ,
\end{equation}
\begin{equation} \label{TemPhiPhi}
T^{\textrm{EM}}_{\phi \phi} = \frac{Q^{2}}{8 \pi} \hspace{0.5 mm} r\left(u, v\right)^{-2} \hspace{0.5 mm} \sin^{2}\left(\theta\right) .
\end{equation}
All other components of $T^{\textrm{EM}}_{\alpha \beta}$ are zero. The total stress-energy tensor $T_{\alpha \beta}$ is the sum of the electromagnetic and scalar field contributions.
\begin{equation} \label{Ttotal}
T_{\alpha \beta} = T^{\textrm{EM}}_{\alpha \beta} + T^{s}_{\alpha \beta}
\end{equation}
From Eqn. \ref{elldef}, we know that the only non-zero component of the null vector field $\ell^{\alpha}$ is $\ell^{v}$. Thus, the null energy condition (Eqn. \ref{nullenergycondn}) is trivially satisfied:
\begin{align}
T_{\alpha \beta} \hspace{0.5 mm} \ell^{\alpha} \hspace{0.5 mm} \ell^{\beta} & = T_{v v} \hspace{0.5 mm} \left(\ell^{v}\right)^{2} \\  \label{nullenergycondnproofbegin}
& = \frac{1}{8 \pi} \hspace{0.5 mm} \left(\phi_{, v}\right)^{2} \hspace{0.5 mm} F\left(u, v\right)^{-2} \\
& \geq 0 \label{nullenergycondnproofend}
\end{align}
Using Eqns. \ref{Tscalar}-\ref{Ttotal}, we may rewrite the Einstein field equations (Eqns. \ref{maineineqn}-\ref{reqnv}) as
\begin{equation} \label{ruvT}
r_{, u v} + \frac{r_{, u} \hspace{0.5 mm} r_{, v}}{r} + \frac{F}{4 r} = 4 \pi \hspace{0.5 mm} r \hspace{0.5 mm} T_{u v} ,    
\end{equation}
\begin{equation} \label{FuvT}
F_{, u v} - \frac{F_{, u} \hspace{0.5 mm} F_{, v}}{F} - \frac{2F}{r^{2}} \hspace{0.5 mm} r_{, u} \hspace{0.5 mm} r_{, v} - \frac{F^{2}}{2 r^{2}} = -\frac{8 \pi \hspace{0.5 mm} F^{2}}{r^{2}} \hspace{0.5 mm} T_{\theta \theta} ,
\end{equation}
\begin{equation} \label{ruuT}
r_{, u u} - \left(\ln F\right)_{, u} \hspace{0.5 mm} r_{, u} = - 4 \pi \hspace{0.5 mm} r \hspace{0.5 mm} T_{u u} ,
\end{equation}
\begin{equation} \label{rvvT}
r_{, v v} - \left(\ln F\right)_{, v} \hspace{0.5 mm} r_{, v} = - 4 \pi \hspace{0.5 mm} r \hspace{0.5 mm} T_{v v}.
\end{equation}
Let us introduce the effective stress-energy tensor $T^{\textrm{eff}}_{\mu \nu}$:
\begin{equation}
T^{\textrm{eff}}_{\alpha \beta} = T_{\alpha \beta} - \frac{\Lambda}{8 \pi} \hspace{0.5 mm} g_{\alpha \beta}
\end{equation}
The tensor $T^{\textrm{eff}}_{\mu \nu}$ incorporates the effects of the cosmological constant $\Lambda$. To account for $\Lambda$ in the Einstein field equations, we should replace $T_{\alpha \beta}$ with $T^{\textrm{eff}}_{\alpha \beta}$ in Eqns. \ref{ruvT}-\ref{rvvT}. We may then write the Einstein field equations as 
\begin{equation} \label{ruvTLambda}
r_{, u v} + \frac{r_{, u} \hspace{0.5 mm} r_{, v}}{r} + \frac{F}{4 r} = 4 \pi \hspace{0.5 mm} r \hspace{0.5 mm} T_{u v} - \frac{\Lambda \hspace{0.5 mm} r}{2} \hspace{0.5 mm} g_{u v} ,    
\end{equation}
\begin{equation} \label{FuvTLambda}
F_{, u v} - \frac{F_{, u} \hspace{0.5 mm} F_{, v}}{F} - \frac{2F}{r^{2}} \hspace{0.5 mm} r_{, u} \hspace{0.5 mm} r_{, v} - \frac{F^{2}}{2 r^{2}} = -\frac{8 \pi \hspace{0.5 mm} F^{2}}{r^{2}} \hspace{0.5 mm} T_{\theta \theta} + \frac{\Lambda \hspace{0.5 mm} F^{2}}{r^{2}} \hspace{0.5 mm} g_{\theta \theta} ,
\end{equation}
\begin{equation} \label{ruuTLambda}
r_{, u u} - \left(\ln F\right)_{, u} \hspace{0.5 mm} r_{, u} = - 4 \pi \hspace{0.5 mm} r \hspace{0.5 mm} T_{u u} ,
\end{equation}
\begin{equation} \label{rvvTLambda}
r_{, v v} - \left(\ln F\right)_{, v} \hspace{0.5 mm} r_{, v} = - 4 \pi \hspace{0.5 mm} r \hspace{0.5 mm} T_{v v}.
\end{equation}
Because $g_{u u}$ and $g_{v v}$ are both zero, there is no cosmological constant term in Eqns. \ref{ruuTLambda} and \ref{rvvTLambda}. From Eqns. \ref{Tscalar}-\ref{Ttotal}, we know all the components of the stress-energy tensor $T_{\alpha \beta}$, which does not include the effects of $\Lambda$. After plugging these expressions for $T_{\alpha \beta}$ into Eqns. \ref{ruvTLambda}-\ref{rvvTLambda}, we obtain
\begin{equation}
r_{, u v} = -\frac{r_{, u} \hspace{0.5 mm} r_{, v}}{r} - \frac{F}{4 r} \hspace{0.5 mm} \left(1 - \frac{Q^{2}}{r^{2}}\right) + \frac{\Lambda}{4} \hspace{0.5mm} r \hspace{0.5 mm} F ,    
\end{equation}
\begin{equation}
F_{, u v} = \frac{F_{, u} \hspace{0.5 mm} F_{, v}}{F} + \frac{2F}{r^{2}} \hspace{0.5 mm} r_{, u} \hspace{0.5 mm} r_{, v} + \frac{F^{2}}{2 r^{2}} \hspace{0.5 mm} \left(1 - \frac{2 Q^{2}}{r^{2}}\right) - 2 F \hspace{0.5 mm} \phi_{, u} \hspace{0.5 mm} \phi_{, v} + \Lambda \hspace{0.5 mm} F^{2} ,
\end{equation}
\begin{equation}
r_{, u u} - \left(\ln F\right)_{, u} \hspace{0.5 mm} r_{, u} + r \hspace{0.5 mm} \left(\phi_{, u}\right)^{2} = 0 ,
\end{equation}
\begin{equation}
r_{, v v} - \left(\ln F\right)_{, v} \hspace{0.5 mm} r_{, v} + r \hspace{0.5 mm} \left(\phi_{, v}\right)^{2} = 0 .
\end{equation}

\section*{Appendix C: Nonlinear Electrodynamics and Wormhole Collapse}
In this section, we generalize the analysis of the previous sections to theories of modified gravity. Of course, it is not practical to address all theories of modified gravity in this section. Thus, we focus on models that couple general relativity to non-linear electrodynamics. As before, we assume spherical symmetry and we use double null coordinates. Thus, in general double null coordinates $\mathcal{U}$ and $\mathcal{V}$, the metric takes the form given in Eqn. \ref{generalsphericallysymmetricmetricuv}.

\subsection*{Wormhole Collapse in a Bardeen Black Hole}
The Reissner-Nordstr\"om solution arises from general relativity coupled to a standard, Maxwellian electromagnetic field. However, it is possible to couple general relativity to non-linear theories of electrodynamics. We focus on a well-known black hole solution in general relativity coupled to non-linear electrodynamics: the Bardeen black hole. 
\vspace{2 mm}
\newline Let $F_{\mu \nu}$ be the electromagnetic field strength tensor. We define the scalar EM field strength as
\begin{equation} \label{Fdef}
F = \frac{1}{4} \hspace{0.5 mm} F_{\mu \nu} \hspace{0.5 mm} F^{\mu \nu} .
\end{equation}
For the Bardeen model, we have the following expression for $F$ \cite{AYONBEATO2000149}:
\begin{equation} \label{Fexpression}
F = \frac{g^{2}}{2 r^{4}} .
\end{equation}
In the Bardeen model, the center of the black hole has mass $m$ and a magnetic charge $g$. The Lagrangian $\mathcal{L}_{E M}$ for the electromagnetic field depends on these parameters. We have the following expression for $\mathcal{L}_{E M}$ \cite{AYONBEATO2000149}:
\begin{equation} \label{bardeenlagrangian}
\mathcal{L}_{E M} = \frac{3 m}{\lvert g \rvert ^{3}} \hspace{0.5 mm} \left(\frac{\sqrt{2 g^{2} \hspace{0.5 mm} F}}{1 + \sqrt{2 g^{2} \hspace{0.5 mm} F}}\right)^{\frac{5}{2}} .
\end{equation}
We define the function $\mathcal{L}_{E M}^{\prime}$ as
\begin{align} \label{LEMprimedef}
\mathcal{L}_{E M}^{\prime} & = \frac{\partial \mathcal{L}_{E M}}{\partial F} \\
& = \frac{15 m}{2 \lvert g \rvert} \hspace{0.5 mm} \left(\frac{\sqrt{2 g^{2} \hspace{0.5 mm} F}}{1 + \sqrt{2 g^{2} \hspace{0.5 mm} F}}\right)^{\frac{3}{2}} \hspace{0.5 mm} \frac{1}{\sqrt{2 g^{2} \hspace{0.5 mm} F} \hspace{0.5 mm} \left(1 + \sqrt{2 g^{2} \hspace{0.5 mm} F}\right)^{2}} \label{LEMprimeexpression} .
\end{align}
From Eqns. \ref{Fexpression} and \ref{LEMprimeexpression}, we know that $\mathcal{L}_{E M}^{\prime}$ is non-negative for the Bardeen solution. We may write the electromagnetic stress-energy tensor as \cite{AYONBEATO2000149}
\begin{equation} \label{generalEMTmunu}
T^{E M}_{\mu \nu} = 2 \mathcal{L}_{E M}^{\prime} \hspace{0.5 mm} F_{\mu \lambda} \hspace{0.5 mm} F_{\nu}^{\lambda} - 2 g_{\mu \nu} \hspace{0.5 mm} \mathcal{L}_{E M} .
\end{equation}
Let the null vector field $\ell^{\mu}$ be defined by Eqn. \ref{elldefmain}. It is easy to verify that Eqn. \ref{generalEMTmunu} satisfies the null energy condition for all possible configurations of the electromagnetic field. In symbols,
\begin{align}
T^{EM}_{\mu \nu} \hspace{0.5 mm} \ell^{\mu} \hspace{0.5 mm} \ell^{\nu} & = 2 \mathcal{L}_{E M}^{\prime} \hspace{0.5 mm} F_{\mu \lambda} \ell^{\mu} \hspace{0.5 mm} F_{\nu}^{\lambda} \ell^{\nu} - 2 g_{\mu \nu} \hspace{0.5 mm} \ell^{\mu} \ell^{\nu} \hspace{0.5 mm} \mathcal{L}_{E M} \\
& = 2 \mathcal{L}_{E M}^{\prime} \hspace{0.5 mm} \hspace{0.5 mm} \left(F_{\mu \lambda} \ell^{\mu}\right)^{2} \\
& \geq 0 .
\end{align}
Let us define the function $f_{B}\left(r\right)$ as
\begin{equation}
f_{B}\left(r\right) = 1 - \frac{2 m \hspace{0.5 mm} r^{2}}{\left(r^{2} + g^{2}\right)^{\frac{3}{2}}} .
\end{equation}
In spherical coordinates, the metric for a Bardeen black hole takes the form \cite{AYONBEATO2000149}
\begin{equation}
ds^{2} = -f_{B}\left(r\right) \hspace{0.5 mm} dt^{2} + f\left(r\right)^{-1} \hspace{0.5 mm} dr^{2} + r^{2} \hspace{0.5 mm} d \Omega^{2} .
\end{equation}
Just like RN black holes, Bardeen black holes may be classified as sub-extremal or extremal. (Neither the super-extremal RN space-time nor the super-extremal Bardeen space-time contain any horizons.) For simplicity, we focus on sub-extremal Bardeen black holes, which satisfy the condition \cite{AYONBEATO2000149}
\begin{equation} \label{subextremebardeen}
g^{2} < \frac{16}{27} \hspace{0.5 mm} m^{2} .
\end{equation}
If Eqn. \ref{subextremebardeen} is satisfied, the Bardeen solution possesses an outer event horizon and an inner Cauchy horizon. After maximal extension (in the absence of perturbations), the sub-extremal Bardeen solution exhibits a wormhole similar to the wormhole that appears in the maximally-extended RN space-time \cite{AYONBEATO2000149}. To describe the Cauchy horizon and the space-time on either side of it, we adopt the coordinate system defined in Subsection \ref{StandardCoordinateSystem}. 
\vspace{2 mm}
\newline Let us define the Schwarzschild mass function as \cite{PhysRevD.103.124027}
\begin{equation}
M\left(u, V\right) = \frac{1}{2} r\left(u, V\right) \hspace{0.5 mm} \left(1 + 4 F\left(u, V\right)^{-1} \hspace{0.5 mm} r_{, u}\left(u, V\right) \hspace{0.5 mm} r_{, V}\left(u, V\right)\right) .
\end{equation}
Everywhere along the Cauchy horizon, the mass function $M\left(u, V\right)$ blows up \cite{PhysRevD.103.124027}. In symbols,
\begin{equation} \label{massinflationbardeen}
\lim_{V \to 0} M\left(u, V\right) = + \infty .
\end{equation}
Let us assume the conditions in Subsection \ref{regconds} are satisfied, along with Eqn. \ref{rurcond}. Then, according to Theorem \ref{thmcol},
\begin{equation} \label{thmcolconclusion1}
r_{, u}\left(u, V = 0\right) > 0 \textrm{ for all } u < u_{1} .
\end{equation}
In Subsection \ref{regconds}, we assumed that $r_{, u}\left(u, V\right)$ is continuously differentiable (and thus finite) at the Cauchy horizon. Thus, in order for Eqns. \ref{massinflationbardeen} and \ref{thmcolconclusion1} to simultaneously be true, the following must be true:
\begin{equation}
\lim_{V \to 0} r_{, V}\left(u, V\right) = + \infty .
\end{equation}
Thus, all the preconditions for Theorem \ref{thmposrv} are satisfied. This implies that the wormhole in the maximally-extended Bardeen solution collapses, and it is impossible for an observer to pass through it.

\subsection*{Generalization to Other Models of Nonlinear Electrodynamics}
Of course, Eqn. \ref{bardeenlagrangian} is not the only possible expression for $\mathcal{L}_{E M}$. Indeed, it is possible to couple general relativity to electrodynamics regardless of the expression for any $\mathcal{L}_{E M}$. Some of these models exhibit solutions similar to the Bardeen black hole \cite{PhysRevLett.80.5056, AYONBEATO199925, Rodrigues_2018}. In the sub-extremal case, each of these space-times contains an event horizon and an inner Cauchy horizon. After maximal extension, these space-times contain wormholes and parallel universe regions.
\vspace{2 mm}
\newline Let $F$ be defined as in Eqn. \ref{Fdef}. For a general model of non-linear electrodynamics, we may express the electromagnetic Lagrangian $\mathcal{L}_{E M}$ as a function of the scalar $F$. In symbols,
\begin{equation}
\mathcal{L}_{E M} = \mathcal{L}_{E M}\left(F\right) .
\end{equation}
Let us consider general relativity coupled to a model of non-linear electrodynamics with Lagrangian $\mathcal{L}_{E M}\left(F\right)$. Furthermore, let us assume that this model has a spherically-symmetric black hole solution with an event horizon, a Cauchy horizon, and a wormhole region. If the conditions in Subsections \ref{regconds} and \ref{specificconds} are true, we may complete the proof of Theorem \ref{thmposrv} only if the null energy condition holds. 
\vspace{2 mm}
\newline In Eqns. \ref{nullenergycondnproofbegin}-\ref{nullenergycondnproofend}, we showed that a massless Klein-Gordon scalar field satisfies the null energy condition. Thus, we may assume the null energy condition holds for the total stress-energy tensor if it holds for the EM stress-energy tensor. The electromagnetic stress-energy tensor is still
\begin{equation}
T^{E M}_{\mu \nu} = 2 \mathcal{L}_{E M}^{\prime} \hspace{0.5 mm} F_{\mu \lambda} \hspace{0.5 mm} F_{\nu}^{\lambda} - 2 g_{\mu \nu} \hspace{0.5 mm} \mathcal{L}_{E M} .
\end{equation}
Let $\mathcal{L}_{E M}^{\prime}\left(F\right)$ be defined as in Eqn. \ref{LEMprimedef}. As long as $\mathcal{L}_{E M}^{\prime} \geq 0$, the EM stress-energy tensor satisfies the null energy condition. In symbols,
\begin{equation}
T^{EM}_{\mu \nu} \hspace{0.5 mm} \ell^{\mu} \hspace{0.5 mm} \ell^{\nu} \geq 0 .
\end{equation}
Thus, as long as $\mathcal{L}_{E M}^{\prime} \geq 0$, all the sufficient conditions for Theorem \ref{thmposrv} are met. Thus, the wormhole inside the Cauchy horizon collapses, and it is impossible for an observer to reach the parallel universe regions of the unperturbed maximally-extended space-time.

\printbibliography

\end{document}